\documentclass[%
 reprint,prb,
groupedaddress,
 amsmath,amssymb,
 aps,
]{revtex4-2}

\usepackage{xcolor}
\usepackage{graphicx}
\usepackage{dcolumn}
\usepackage{siunitx}
\usepackage{bm}
\usepackage{hyperref}
\usepackage{float}
\usepackage[english]{babel}

\begin{document}

\title{Optical studies of negative \texorpdfstring{$g$}{g}-factor dopants in an antiferromagnet: Nd:GdVO\texorpdfstring{$_4$}{₄}}

\author{Lachlan W. Cronin}
\author{Masaya Hiraishi}
\author{Luke S. Trainor}
\author{Jevon J. Longdell}
\affiliation{Department of Physics, University of Otago, Dunedin, New Zealand}
\affiliation{Dodd-Walls Centre for Photonic and Quantum Technologies, Dunedin, New Zealand}
\email{Jevon.Longdell@otago.ac.nz}

\date{\today}

\begin{abstract}
  Neodymium-doped gadolinium vanadate (Nd:GdVO$_4$) is a popular laser material at room temperature. It has received limited attention for quantum technologies because of the linebroadening effects of disordered gadolinium spins in the host. Here we investigate the optical properties at temperatures below the \qty{2.5}{\K} N\'eel temperature, revealing narrow optical transitions.
  We demonstrate that rare earth ions in magnetic hosts give us a new way of observing the signs of magnetic $g$-factors. The interplay between the exchange and Zeeman interactions means that we see clear evidence of negative $g$-factors. 
  These results provide a detailed spectroscopic foundation for Nd:GdVO$_4$ and expand the understanding of its candidacy as a material for hybrid quantum systems.    
\end{abstract}

\maketitle

\section{\label{sec:intro} Introduction
}
Rare earth ions are an attractive platform for coherent and quantum information processing because they offer long lived coherences for both optical \cite{bottger_effects_2009,equall_ultraslow_1994} and spin transitions \cite{rancic_coherence_2018,zhong_optically_2015,rakonjac_long_2020} in the solid state. Ensembles of rare earth ions are being pursued as quantum memories \cite{ortu_storage_2022,meng_efficient_2026,8l9k-12k2,z6lc-qw2d,stuart_progress_2025},  microwave-optical transducers \cite{PhysRevLett.113.203601,PhysRevA.100.033807,xie_scalable_2025}, 
and single emitters in optical resonators \cite{kindem_control_2020,ourari_indistinguishable_2023,Deshmukh:23}.

In all these investigations, a key consideration is the dephasing effect of the ``magnetic noise" of the other spins in the crystal. For rare-earth ions used in quantum and coherent information processing, host lattices are chosen to minimize magnetic moments. Both electronic and nuclear spins act as sources of magnetic noise, driving decoherence of rare-earth optical and spin transitions \cite{thielOpticalDecoherenceStudies2012, onizhukUnderstandingCentralSpin2024}.

Generally, electron spins in the host crystal have been avoided because of their large magnetic moments, which cause large amounts of dephasing. However, it has been shown recently that electron spins might not be as undesirable as previously thought. The large moment means that it is possible to strongly order them at sufficiently low temperatures.
Investigations of erbium dopants in gadolinium vanadate (GdVO$_4$) showed that coherence times similar to non-magnetic yttrium vanadate were obtained \cite{hiraishiLongOpticalCoherence2025}. For both yttrium vanadate (YVO$_4$) and GdVO$_4$ at low temperatures the dominant dephasing contribution was from the $^{51}$V nuclear spins. 

Ideally, a host would be free from both electron and nuclear spins. However, due to an unfortunate coincidence of nuclear and atomic physics effects, there are no stable isotopes that form trivalent rare-earth-like ions that are free of both nuclear and electronic spins. 
This reality has caused an increased focus on materials like calcium tungstate (CaWO$_4$) which has a very low spin density at the cost of greater misfit between the rare earth ion and the host ion it is substituting \cite{tiranovSubsecondSpinLifetimelimited2025,ledantecTwentythreemillisecondElectronSpin2021}.

Magnetic hosts offer an alternative route to nuclear-spin-free host crystals. Gadolinium ions are an attractive option: they have no optical transitions in the visible or infrared. Compared to yttrium, gadolinium has a lower nuclear spin density \cite{kanaiGeneralizedScalingSpin2022} and the spin free isotopes have a combined abundance of nearly 70\%. 

In addition to long coherence times, strong coherent coupling between the erbium dopants and magnetic resonances in the host gadolinium spins was also observed in \textcite{hiraishiLongOpticalCoherence2025}, suggesting a potential route to better bandwidths and efficiencies for microwave to optical transduction. The gadolinium electron spins in GdVO$_4$ couple strongly to microwave cavities \cite{evertsUltrastrongCouplingMicrowave2020} due to the large densities of ions, meaning transduction through magnons in Nd:GdVO$_4$ is alluring.

With antiferromagnetically ordered systems, such as the one we investigate here, there are strong internal exchange fields but no macroscopic magnetisation. This means that there are microwave frequency spin excitations with the potential for long coherence times without macroscopic magnetic fields. This is attractive particularly when working with superconducting devices that are adversely affected by magnetic fields \cite{krantzQuantumEngineersGuide2019a}. 

This paper investigates the optical spectroscopy of Nd$^{3+}$ dopants in magnetically ordered gadolinium vanadate, building on our earlier work with erbium dopants in the same system \cite{hiraishiLongOpticalCoherence2025}. 

Because the Nd$^{3+}$ has both exchange interactions with its neighboring spins and  Zeeman interactions with the (both internal and applied) magnetic field, it also presents a new, unique way of observing the signs of the magnetic $g$-factors.

\section{What is a negative \texorpdfstring{$g$}{g}-factor?}

Usually observed spectra are insensitive to the signs of the magnetic $g$-factors, and are often assumed to be positive. In Nd:GdVO$_4$ the signs of the $g$-factors do affect the spectra.

From the success of semi-empirical crystal field models for rare-earth dopants, we can calculate rare earth ion properties using real spin operators $\bm{L}$ and $\bm{S}$. In this discussion, we will assume that the $z$ axis is a principal axis. The $g$-factor is proportional to the magnetic moment that an ion will have in an applied magnetic field. For the two states of a Kramers doublet, with a $z$ quantisation axis, $\lvert a\rangle,\lvert b\rangle$, the $g$-factor can be taken to be
\begin{equation}
    g_{zz} = \langle a\rvert L_z + g_sS_z\lvert a\rangle  -\langle b\rvert L_z + g_sS_z\lvert b\rangle,
\end{equation}
where $g_s\approx 2$ is the bare electron spin $g$-factor. 
In this definition, however, there is a sign ambiguity as we have not defined which state of the doublet is $\lvert a\rangle$, and which is $\lvert b\rangle$. Choosing states $\lvert\uparrow\rangle,\lvert\downarrow\rangle$, such that $\langle \uparrow \rvert S_z\lvert \uparrow\rangle >0 >\langle \downarrow \rvert S_z\lvert\downarrow\rangle $, then we can uniquely define
\begin{equation}
    g_{zz} = \langle \uparrow\rvert L_z + 2S_z\lvert \uparrow\rangle  -\langle \downarrow\rvert L_z + 2S_z\lvert \downarrow\rangle,
\end{equation}
and the sign of the $g$-factor becomes clear. It is positive in the usual case that the magnetic moment $\langle\mu_i\rangle=-\mu_B\langle L_i + 2S_i\rangle$ opposes the spin $\langle S_i\rangle$,
but can be negative if they are in the same direction; i.e. the orbital angular momentum is large and counteracts the spin. In our antiferromagnetic system, the spin dominates the ordering at low applied field, whereas the magnetic moment takes over in high fields.
Therefore, by comparing low- and high-field optical spectra, the $g$-factor signs become clear experimentally.

Previously, the sign of $g$ has been observed in other ways such as circularly polarized electron paramagnetic resonance (EPR) \cite{hutchisonParamagneticResonanceAbsorption1960}. If we assume a uniaxial point group symmetry, such as the $D_{2d}$ symmetry we have here, then in the $\lvert\uparrow\rangle,\lvert\downarrow\rangle$ basis, the Zeeman Hamiltonian takes the form
\begin{equation}
    H_Z=\mu_B\left[g_\perp \left(B_x \tilde{S}_x + B_y\tilde{S}_y\right) + g_\parallel B_z\tilde{S}_z\right],
\end{equation}
where the effective spin-\textonehalf{} operators are proportional to the Pauli matrices $\tilde{S}_i=\sigma_i/2$.
Assuming we have a static field along $z$, such that the eigenstates are $\lvert\uparrow\rangle,\lvert\downarrow\rangle$ and they are split in frequency,
the transition operator for circularly polarized (microwave) photons of a particular polarization is proportional to
\begin{equation}
    L_x+2S_x+i(L_y+2S_y) \rightarrow g_\perp(\tilde{S}_x+i\tilde{S}_y) = g_\perp\tilde{S}_+.
\end{equation}

Therefore, that circular polarization is a raising operator in the $\lvert\uparrow\rangle,\lvert\downarrow\rangle$ basis, but the sign of $g_\parallel$ dictates which of those two states is the ground state. In either case the ground state has its magnetic moment parallel to the applied field. Therefore, the sign of $g_\parallel$ determines which circular polarization will raise the ions from their ground state, and if $g_\parallel<0$, the magnetic moments of the ions are said to precess in the opposite direction to a free electron \cite{rignyResonanceParamagnetiqueDans1967}. More generally it is the sign of $\det(\bm{g})$ which determines precession sense \cite{pryceSign$g$Magnetic1959}.

\section{Sample information and methodology}

GdVO$_4$ has the same zircon structure as YVO$_4$ crystallising in the tetragonal space group I4$_1$/amd. The Gd$^{3+}$ sites have D$_{2d}$ point symmetry \cite{mullicaStructuralInvestigationsSeveral1996}  and are coupled to their four nearest neighbors through oxygen-mediated superexchange interactions that stabilise long-range magnetic order. 

Gd$^{3+}$ has a half-filled $4f$ shell, giving rise to a $^8\text{S}_{7/2}$ ground state. Consequently, its magnetic moment is almost entirely due to electron spin ($L\approx0$). 

Below the N\'eel temperature of \SI{2.495}{\kelvin}, the Gd$^{3+}$ moments order antiferromagnetically in a $G_z$-type structure, where each spin is oriented antiparallel to all nearest neighbors, forming two interpenetrating sublattices with moments aligned along the crystallographic $c$-axis \cite{cashion1970, andreicieftimieMorphologyGdVO4Crystal2020}.

Building on previous work done in the group on Er$^{3+}$:GdVO$_4$ \cite{hiraishiLongOpticalCoherence2025}, Nd$^{3+}$ is employed as the dopant ion, whose ${}^4\text{I}_{9/2}\rightarrow{}^4\text{F}_{3/2}$ transition exhibits an unusually strong $4f -4f$ oscillator strength \cite{sunRecentProgressDeveloping2002}. This transition occurs around \SI{880}{\nano\metre}, a technologically accessible wavelength for diode-laser-based spectroscopy and potential device integration. In related zircon-type hosts such as YVO$_4$, this transition is strongly $\pi$-polarized with respect to the crystallographic $c$-axis \cite{sunRecentProgressDeveloping2002}, motivating the use of polarization-selective spectroscopy to resolve site- and phase-dependent spectral features.

Nd$^{3+}$ has an odd number of electrons and is therefore a Kramers ion. In the absence of time-reversal-symmetry-breaking interaction, its electronic states will occur as at least doubly degenerate Kramers doublets. The free-ion ${}^4\text{F}_{3/2}$ manifold has $J=3/2$ and is four-fold degenerate; the crystal field lifts this degeneracy into two Kramers doublets. 
As is characteristic of GdVO$_4$ and YVO$_4$ the crystal field splitting of the excited ${}^4\text{F}_{3/2}$ multiplet is small, about \qty{150}{\GHz} in this case. 

In the antiferromagnetically ordered phase of GdVO$_4$ exchange coupling between Nd$^{3+}$ and the ordered Gd$^{3+}$ moments breaks Kramers degeneracy, splitting the two doublets into four nondegenerate levels. In this ordered phase, there are two magnetically nonequivalent Nd$^{3+}$ sites: spin up and spin down. At zero field, their levels are degenerate, but in an applied magnetic field the Zeeman effect tunes each orientation differently.
Therefore, each orientation has a distinct set of four ${}^4\text{F}_{3/2}$ states, giving eight possible transitions.
Note that at millikelvin temperatures, thermal population is restricted to the lowest level of the ${}^4\text{I}_{9/2}(Z_1)$ doublet.

\begin{figure}
    \centering
    \includegraphics[width=0.9\columnwidth]{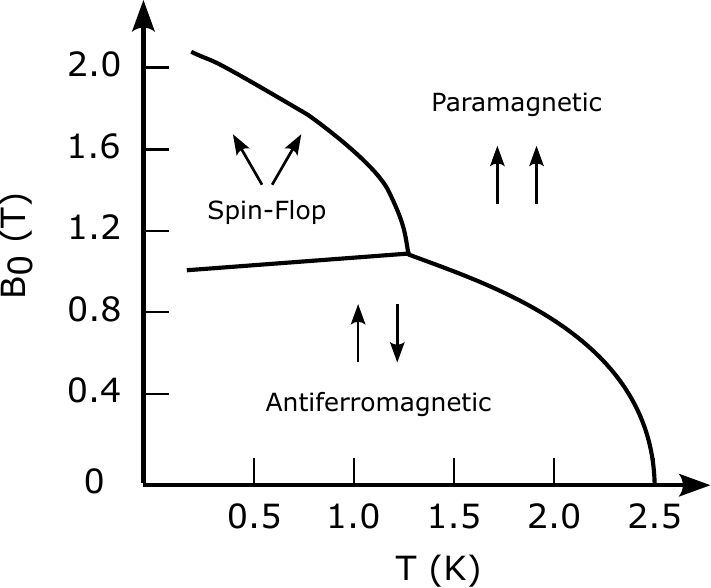}
    \caption{Magnetic phase diagram of GdVO$_4$ as a function of temperature and external field applied along the crystallographic $c$-axis. Arrows indicate the relative orientations of spin alignments in the paramagnetic (PM), antiferromagnetic (AFM), and spin-flop (SF) phases. The phase diagram is adapted from \cite{EvertsJonathan2020IoFC, moonMagneticStructureEr2O31967a, freemanTheoreticalInvestigationMagnetic1962a}}
    \label{fig:phases}
\end{figure}

Over the magnetic-field range investigated, the GdVO$_4$ host lattice
has three distinct magnetic phases: antiferromagnetic (AFM), spin-flop (SF), and paramagnetic (PM), as shown in Fig.~\ref{fig:phases}. Each phase provides a different magnetic environment for the Nd$^{3+}$ ions. In the low-field AFM phase, the two Gd$^{3+}$ sublattices are antialigned, producing the two nonequivalent magnetic environments of the Nd$^{3+}$ dopant ions. As the applied field increases, Zeeman energy competes with antiferromagnetic exchange, causing the sublattice moments to cant and drive a spin-flop transition. At higher fields, the moments progressively align with the external field and, in the paramagnetic regime, the sublattice distinction is lost, rendering the Nd$^{3+}$ sites magnetically equivalent. In this work, we probe the magnetic-field-dependent optical response of Nd:GdVO$_4$ across these phases using polarization-resolved absorption spectroscopy. By combining the measured spectra with crystal-field modelling, we assign the observed transitions and map the evolution of the ${}^4\text{F}_{3/2}$ energy manifold with magnetic field.

    The nominally undoped GdVO$_4$ sample contained an estimated 10--100\,ppm of Nd$^{3+}$, estimated from its coupling strength to a microwave cavity, with trace Er$^{3+}$ impurities also present \cite{MasayaThesis}.

    The sample used was purchased as undoped GdVO$_4$ and machined into a cylinder of diameter \SI{2.5}{\milli\meter} and length \SI{4.2}{\milli\meter}. This sample was mounted in, and cooled using a Bluefors LD-250 dilution refrigerator.
    
    Based on prior runs we approximate the sample temperature as $\approx$ \SI{70}{\milli\kelvin}. The crystal was positioned within the bore of a \SI{3}{\tesla} tuneable superconducting magnet, with the crystal’s $c$-axis approximately aligned to the external field. Cross-polarization measurements indicated a misalignment of roughly \ang{10}. This configuration enabled optical propagation along one of the two-fold degenerate $a$-axes, allowing experimental selection of light polarization either perpendicular ($\sigma$) or parallel ($\pi$) to the $c$-axis.

    Absorption spectroscopy was performed using a Ti:sapphire laser (M~Squared SolsTiS~PI), swept from \SI{878.35}{\nano\meter} to \SI{879.70}{\nano\meter} (\SIrange{341300}{340800}{\giga\hertz}) in $\approx\SI{30}{\giga\hertz}$ segments. A wavemeter (HighFinesse WS7) recorded the absolute optical frequency at the beginning and end of each sweep.

\begin{figure}
    \centering
    \includegraphics[width=0.95\linewidth]{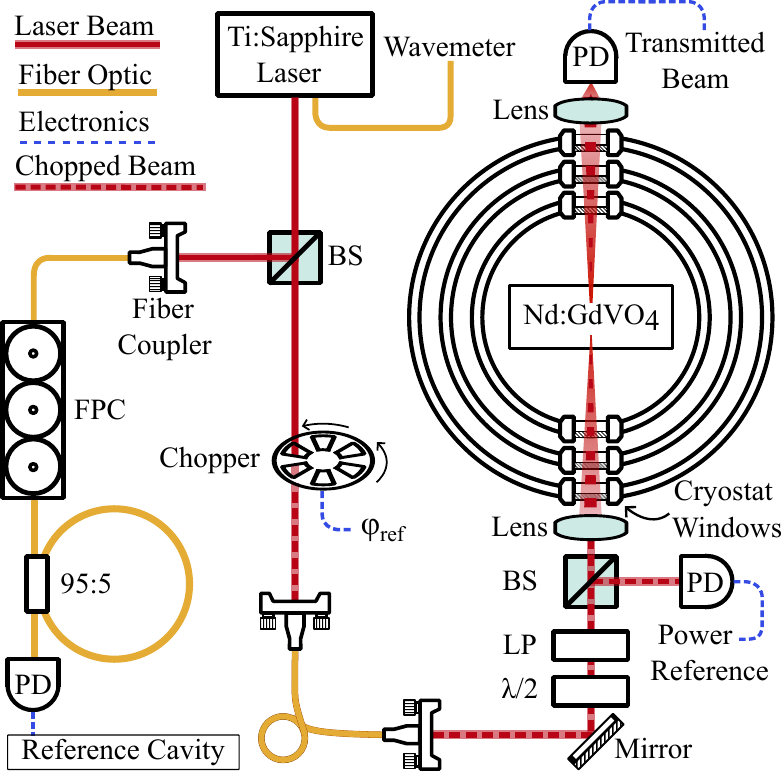}
    \caption{Diagram of the experimental setup used for absorption spectroscopy of Nd:GdVO$_4$. A tuneable laser is directed through the crystal mounted in a cryostat, with the magnetic field applied along the crystallographic $c$-axis. Transmitted light is detected using a photodiode. Some lenses, mirrors, and power attenuating components have been omitted for clarity. The power reference and the transmitted beam were fed to two separate lock-in amplifiers, along with $\varphi_\textrm{ref}$. BS, beamsplitter; FPC, fiber polarization controller; LP, linear polarizer; PD, photodetector.}
    \label{fig:setup}
\end{figure}

    The experimental setup (Fig. \ref{fig:setup}) began with attenuation of the laser power. 
    To avoid heating of the sample $\approx$ \SI{20}{\micro\watt} was used for the absorption measurements. A small amount of light was diverted with a beam splitter (BS), and fiber-coupled to a temperature-stabilised reference cavity made of a 95:5 fiber splitter
    with a known free spectral range (FSR) of \SI{151.7\pm0.1}{\mega\hertz}.
    The reference cavity allowed the nonlinear frequency sweep of the laser to be corrected for.

    The main beam was modulated using a mechanical chopper to allow lock-in detection. The beam was then fiber coupled and the light was routed to a small breadboard attached to the dilution refrigerator. It was returned to free space, passed through a half-wave plate ($\lambda$/2) and linear polarizer to select $\sigma$ or $\pi$ polarization, and then through a PBS, which partially diverted the beam to a photodetector providing a power reference.

    The remaining beam was focused on the sample through four sets of optical windows on the cryostat.
    
    The transmitted light was collected and detected on a second photodetector. Final absorption spectra were obtained by taking the ratio of the transmitted signal to the power reference signal, with each signal separately undergoing lock-in detection before combination. 
 
\subsection{Mean field model}

    To assist in the assignment of the observed spectral transitions, an expanded crystal-field-theory (CFT) model, containing mean field and exchange, of the Nd$^{3+}$ ions embedded into the host environment was constructed. 

    The Hamiltonian used to address the system was formulated as follows:
    \begin{equation}
    \label{eq:fullham}
        H = H_{\mathrm{Nd}} + H_{\mathrm{ex}} + H_{\mathrm{Gd}}^{\mathrm{mf}} + H_{B},
    \end{equation}
    This total Hamiltonian is a composite of four physically distinct contributions. $H_\mathrm{Nd}$ is the Nd$^{3+}$ crystal-field Hamiltonian, $H_{\mathrm{ex}}$ captures the exchange between the Nd$^{3+}$ ion and its four nearest neighbor Gd$^{3+}$ ions, $H_\mathrm{Gd}^\mathrm{mf}$ encapsulates the effect of the antiferromagnetic mean field on the Gd$^{3+}$ ions, and $H_{B}$ is the Zeeman effect acting on the Nd$^{3+}$ and Gd$^{3+}$ ions.

    The D$_{2d}$ site symmetry determines the set of real crystal field parameters used here \cite{newmanPointSymmetry2000}. By fitting the parameters of the Hamiltonian to experimental transition energies, the model provides both a quantitative description of the observed spectra and a basis for line assignment.

    The Nd$^{3+}$ contribution is written in the general form
    \begin{equation}
        H_{\mathrm{Nd}} = \sum_{p} \lambda_p \hat{O}_p,
    \end{equation}
    where $\hat{O}_p$ are fixed operator matrices acting in the Nd$^{3+}$ $4f$ basis and $\lambda_p$ are parameters including the Slater integrals $F^k$, the spin--orbit coupling constant $\zeta$, and the crystal-field coefficients $B^{(k)}_q$. Only terms permitted by the site symmetry were included.

    The full $4f^3$ Hamiltonian was constructed in the $|\alpha LSJ, m_J\rangle$ basis for Nd$^{3+}$, 
    with matrices constructed using the Python package \texttt{dieke} \cite{dieke}, giving a total dimensionality of 364. 

    The Nd--Gd exchange is modelled by a Heisenberg exchange,
    \begin{align}
        H_\mathrm{ex} &= -2J\bm{S}^\mathrm{Nd}\cdot\sum_{i=1}^z\bm{S}^\mathrm{Gd}_i\\
        &\equiv-2J\bm{S}^\mathrm{Nd}\cdot\bm{S}^\mathrm{Gd}
    \end{align}
    where $i$ labels the $z=4$ nearest neighbor gadolinium ions, $\bm{S}^{Gd}_i$ is the vector of gadolinium spin operators for ion $i$, and $\bm{S}^{Gd}$ is the vector of total gadolinium spin operators.
    Assuming a positive mean field, the lowest two states of the gadolinium ions are $m_S=-7/2,-5/2$.
    We projected this Hamiltonian onto the two $S$-symmetry modes of the gadolinium ions with either no Gd-excitation, i.e. $\lvert-7/2,-7/2,-7/2,-7/2 \rangle$, or a single Gd-excitation $\frac{1}{\sqrt{z}}\sum_{i=1}^z\lvert -7/2,\ldots,-5/2_i,\ldots,-7/2\rangle$.
    In this basis---defining $S_\mathrm{Gd}=7/2$ as the gadolinium ion spin, and $\sigma_i$ as Pauli operators on this two-level subspace---the total gadolinium spin operators become
    \begin{subequations}
    \begin{align}
        S^\mathrm{Gd}_x &= \sqrt{2zS_\mathrm{Gd}}\frac{\sigma_x}2,\\
        S^\mathrm{Gd}_y &= \sqrt{2zS_\mathrm{Gd}}\frac{\sigma_y}2,\\
        S^\mathrm{Gd}_z &=\left(-zS_\mathrm{Gd}+\tfrac12\right)\mathbb{I} + \frac{\sigma_z}2.
    \end{align}
    \end{subequations}

    The Gd$^{3+}$ ions experience a mean field from their other nearest-neighbor gadolinium ions. Assuming a static antiferromagnetic ordering of those ions, the mean-field coupling can be expressed as
    \begin{equation}
        H_{\mathrm{Gd}}^{\mathrm{mf}} 
        = 2\mu_B B_{\mathrm{mf}} S_z^\mathrm{Gd},
    \end{equation}
    where $B_\mathrm{mf}$ is the mean field due to neighboring gadolinium ions as well as anisotropy.  

    The Zeeman interaction was given by
     \begin{equation}
        H_B = \mu_B \bm{B} \cdot
        \Big[
            \bm{L}^\mathrm{Nd} + 2\bm{S}^\mathrm{Nd}
            + 2\bm{S}^\mathrm{Gd}
        \Big],
    \end{equation}
    which accounts for both the orbital and spin contributions of Nd$^{3+}$ and the spin-only contribution from Gd$^{3+}$.

    Due to the low magnitude of $g$-factors for the Nd$^{3+}$ ions, we neglected dipole--dipole interactions between Nd and Gd in the magnetic lattice.

    An important note is that the assumed exchange interaction, coming from the isotropic term of the exchange potential \cite{levyRareEarthIronExchangeInteraction1964}, is not isotropic when projected onto an effective spin Hamiltonian, because it depends on the real spin operators.
    The appropriate projection is
    \begin{equation}
        J \bm{S}^{\mathrm{Nd}} \cdot \bm{S}^\mathrm{Gd}\rightarrow J ( \bm{f} \cdot \tilde{\bm{S}}^{\mathrm{Nd}})\cdot \bm{S}^\mathrm{Gd},
    \end{equation}
    where $\tilde{\bm{S}}^{\mathrm{Nd}}$ is the effective-\textonehalf{} operator acting within the ground state Kramers doublet. The $\bm{f}$-tensor is defined analogously to the $\bm{g}$-tensor, whereas $\bm{g}$ is for $\bm{L}+2\bm{S}$, $\bm{f}$ is for $\bm{S}$, re-emphasising the importance of the $g$-factor signs.
    
    Initial crystal-field parameters were taken from \textcite{andersonInterpretiveCrystalfieldParameters1994c} and free-ion parameters were taken from LaF\textsubscript{3} parameters \cite{carnallSystematicAnalysisSpectra1989}.

    To reduce computational cost during optimization, the crystal-field Hamiltonian was projected onto a truncated Hilbert space spanning the lowest 72 eigenstates. 
    This subspace extends up to the ${}^4\text{F}_{5/2}$ and ${}^2\text{H(2)}_{9/2}$ states so as to capture mixing between those manifolds. Initial parameter fitting was performed within this reduced space, before optimizing using the full Hamiltonian.
    Instead of including a global energy offset as a fitting parameter, often denoted $E_{av}$, we offset all levels such that the lowest energy level was at \qty{0}{\per\cm}.

    optimization was carried out in two stages. First, a coarse fit was performed to establish overall agreement with the ${}^4\text{I},{}^4\text{F}_{3/2},{}^4\text{F}_{5/2}$ energy levels from $H_\mathrm{Nd}$ with experimentally reported zero-field transition energies from the literature \cite{andersonInterpretiveCrystalfieldParameters1994c}, together with transitions resolved in this work. In this stage, free-ion barycenter energies and crystal-field parameters were refined to reproduce the global energy level structure. With the minimization parameters expressed as a vector $\bm{x}$, the coarse fit minimized the objective function
    \begin{equation}
        f_\text{coarse}(\bm{x}) = \hspace{-5mm}\sum_{i\in{}^4\text{I},{}^4\text{F}_{3/2},{}^4\text{F}_{5/2}} \hspace{-5mm} \frac{(E_{i,\text{calc,nonmag.}}(\bm{x}) - E_{i,\mathrm{exp}})^2}{2\sigma_i^2},
    \end{equation}
    where measured transitions from the literature were given an uncertainty, $\sigma_i$, of \qty{4}{\per\cm}, and our measured transitions an uncertainty of \qty{5}{\GHz}.
    
    Second, the misfit was calculated against a fine fit of our well-resolved transitions within the ${}^4\text{F}_{3/2}$ manifold. 
    
    The misfit of these fine transitions to values from Eq.~\eqref{eq:fullham} was calculated as
    \begin{equation}
        f_\text{fine}(\bm{x}) = \hspace{-2mm}\sum_{j\in{}^4\text{F}_{3/2}\otimes\mathrm{Gd}} \hspace{-2mm} \frac{(E_{j,\text{calc,mag.}}(\bm{x}) - E_{j,\mathrm{exp}})^2}{2\sigma_j^2},
    \end{equation}
    where $\sigma_j=\qty{0.2}{\GHz}$.
   The overall fit minimized
    \begin{equation}
        f(\bm{x}) = f_\text{coarse}(\bm{x}) + f_\text{fine}(\bm{x}).
    \end{equation}
    
    All minimization steps used the BFGS quasi-Newton algorithm, with gradients evaluated via forward-mode automatic differentiation \cite{revelsForwardModeAutomaticDifferentiation2016}.
    After minimization, the Hessian matrix $\partial^2 f/\partial x_i\partial x_j$ was calculated using finite differences. For Gaussian random variables, and this definition of the objective function, its inverse is the covariance matrix, from which we found estimates for the parameter uncertainties from the square roots of the diagonal.

    Selection rules for Nd$^{3+}$ optical transitions can be derived from the crystal-field quantum numbers of the levels obtained from the model \cite{marinoEnergyLevelStructure2016}. In the $D_{2d}$ crystal field, $m_J$ is no longer a good quantum number; instead, the appropriate quantum number is the crystal field quantum number $\mu\in\{\pm1/2,\pm3/2\}$. 
    A state with crystal field quantum number $\mu$ is made of components with $m_J=\mu+4n$, where $n$ is any integer. The equivalence of $m_J$ values differing by four comes from the $S_4$ symmetry of the $D_{2d}$ crystal field.
    For the ${}^4\text{I}_{9/2}\rightarrow{}^4\text{F}_{3/2}$ transition, $\Delta J=3$, so magnetic-dipole transitions are forbidden, while forced electric dipole transitions are allowed through odd-parity crystal-field mixing. In this case the electric-dipole selection rules can be simply stated as $\pi$ polarization for $\Delta\mu\in\{\pm2\}$, and $\sigma$ for $\Delta\mu\in\{\pm1,\pm3\}$. Transitions for $\Delta\mu=0$ are not allowed \cite{marinoEnergyLevelStructure2016, Reid2016TheoryOR}.

\begin{figure*}[ht]
    \centering
    \includegraphics[width=\linewidth]{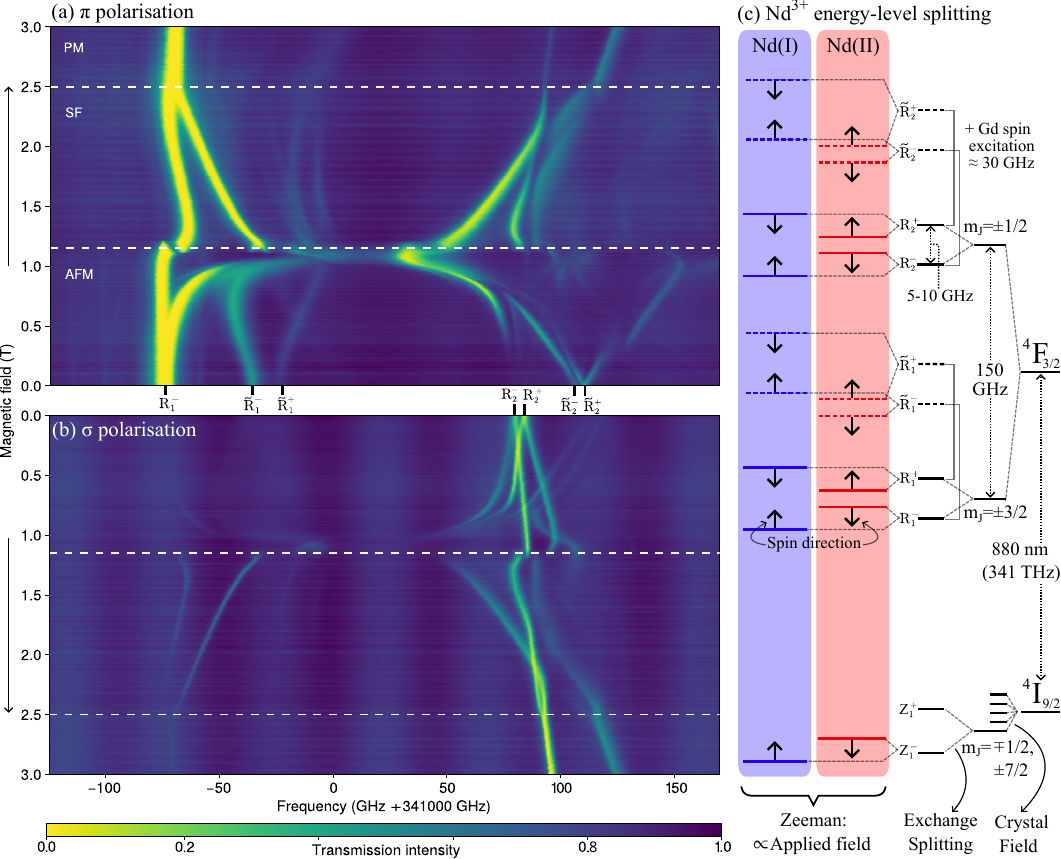}
    \caption[]{
            Plots (a) and (b) display the two-dimensional transmission spectra recorded at base temperature as a function of laser frequency (offset by \SI{341000}{\giga\hertz}) against the applied external magnetic field. The white dotted lines separate the three observed phases. (a) is the transmission data recorded for $\pi$-polarized light ($\bm{E}\parallel c$), and (b) for $\sigma$-polarized light ($\bm{E}\perp c$). Diagram (c) shows an energy-level diagram for transitions between the ground state (Z$_1^-$) and ${}^4\text{F}_{3/2}$ manifold. 
            The ${}^4\text{F}_{3/2}$ doublets are labelled $R_1,R_2$ referencing their ordering in energy. The doublets are split by the exchange interaction and the lower (higher) frequency state at zero applied field is denoted with a superscript $-$ ($+$).
            States which involve a simultaneous Gd spin excitation are denoted with a tilde \~{}.
            }
    \label{fig:spectra}
\end{figure*}

\section{Results} 

\subsection{Experimental spectra}

    Absorption spectroscopy of the ${}^4\text{F}_{3/2}$ manifold of Nd:GdVO$_4$ was performed at $\approx\qty{70}{\milli\K}$ under applied magnetic fields from \SIrange{0}{3}{\tesla}. Spectra were recorded for both $\pi$-polarized light ($\bm{E}\parallel c$) and $\sigma$-polarized light ($\bm{E}\perp c$) over an optical frequency window centerd at \SI{341000}{\giga\hertz}, with an effective span from \SIrange{-150}{+220}{\giga\hertz} relative to this offset. This range encompasses all detectable transitions to the ${}^4\text{F}_{3/2}$ manifold.

    The data are presented as two-dimensional transmission spectra with laser frequency on the horizontal axis and applied magnetic field on the vertical axis (Fig.~\ref{fig:spectra}). The color scale represents the normalized transmitted intensity through the crystal, where decreases in intensity correspond to optical absorptions. 

    The three distinct magnetic phases of GdVO$_4$, corresponding to antiferromagnetic (AFM), spin-flop (SF), and paramagnetic (PM), are evident as the magnetic field is increased. The approximate phase boundaries, indicated by the white dotted lines in Fig.~\ref{fig:spectra}, occur near \SI{1.15}{\tesla} and \SI{2.50}{\tesla}.

    In the low-field AFM regime ($|B| \lesssim \SI{0.8}{\tesla}$), the magnetisation of the Gd-hosts is unchanging, meaning transitions exhibit linear shifts in frequency against applied field. Pairs of branches with opposite slopes are observed, reflecting the presence of two magnetically nonequivalent antiferromagnetic sublattices with opposing local spin orientations. As a result, Zeeman shifts of the Nd$^{3+}$ ions depend on the relative alignment between the applied field and the local exchange field of each sublattice.

    Within the field range \SIrange{1.05}{1.2}{\tesla}, several transitions disappear abruptly as the transition into the SF phase occurs. In this regime, new absorption features emerge with altered field dependence, while some weakly absorbing transitions persist throughout the phase. The strongest transitions gradually converge in frequency with increasing field, reflecting the reorientation of the sublattice magnetisations.

    Figures~\ref{fig:spectra}(a) and (b) show spectra recorded using $\pi$- and $\sigma$-polarized excitation, respectively. At zero applied field, five absorption features are observed in $\pi$ and two in $\sigma$, giving a total of seven resolved zero-field transitions. One additional transition predicted for the ${}^4\text{F}_{3/2}$ manifold is not observed at low fields in either polarization.
    The strongest absorption in the lower-frequency group occurs near \SI{-70}{\giga\hertz} under $\pi$ polarization, while the strongest absorption in the higher-frequency group occurs near \SI{+80}{\giga\hertz} under $\sigma$ polarization. Across all observed lines, full widths at half maximum range from approximately \SIrange{2}{5}{\giga\hertz}.

    The measured properties of all assigned transitions, including center frequency, linewidth, polarization, and field sensitivity, are summarised in Table~\ref{CombinedTable}.
    Profiles of the transitions were modelled using both Gaussian and Lorentzian fits. Lorentzians were found to more accurately replicate the observed line shapes, and were used to find the linewidths.
    
    Field sensitivities were extracted from linear fits to the transition frequencies in the AFM linear regime. The slopes $\mathrm{d}\nu/\mathrm{d}B$ are reported in units of \si{\giga\hertz\per\tesla}, from which effective $g$-factors can be obtained using
    \begin{equation}
         |g| = \frac{h}{\mu_B}\left|\frac{\mathrm{d}\nu}{\mathrm{d}B}\right|.
    \end{equation}

    The zero-field spectrum can be analysed by considering the crystal field splitting transitions by about \SI{150}{\giga\hertz}, breaking them into the $m_J=\pm\tfrac{3}{2}, \,\text{and } \pm \tfrac{1}{2}$ groups. The two groups are at \num{-50}$\pm$\SI{25}{\giga\hertz} (denoted $R_1$, following \cite{diekeSpectraDoublyTriply1963}) and \num{100}$\pm$\SI{25}{\giga\hertz}  (denoted $R_2$). An energy level diagram is shown in Fig.~\ref{fig:spectra}(c) and properties of the transitions are shown in Tab. \ref{CombinedTable}.
        
    In both groups, the respective lowest-frequency branches are the strongest absorption features, which are each observed under orthogonal polarization.
    The higher-frequency lines of each pair exhibit reduced absorption strength. Notably, these weaker transitions are observed under $\pi$ polarization.
    
    Based on their zero-field frequencies, polarization dependence and field evolution, four principal electronic transitions are identified and labelled $R_1^-$, $R_1^+$, $R_2^-$, and $R_2^+$ (see Fig. \ref{fig:spectra}). 
    Here, the superscript indicates the relative frequency ordering within each pair. Cooperative transitions with a Gd spin excitation are denoted with a tilde.

    Additional measurements of the next three higher-energy transitions were performed, revealing transitions at \num{370887}, \num{371860}, and \SI{372125}{\giga\hertz} ($\approx$ \SI{808}{\nano\metre}). Based on comparison with the literature \cite{jensenSpectroscopicCharacterizationLaser1994b,guillot-noelOpticalSpectraCrystal1998}, we included these as the next three doublets in our coarse fit model as the first three states of the mixed ${}^4\text{F}_{5/2}$ and ${}^2\text{H(2)}_{9/2}$ levels.

    \begin{table*}[ht]
    \centering
    \caption[Measured and calculated properties of R$_{1,2}^\pm$ transitions]{Measured and calculated properties of the $R_{1,2}^\pm$ transitions and cooperative transitions with a Gd spin excitation. $\tilde{R}$ indicates cooperative transitions. }
    \begin{tabular}{c|cccccccc}
        \hline
        Transition & Observed$^\dagger$ & Calculated & $m_J$ & Linewidth & polarization & Expected& Field sensitivity & Calculated sensitivity\\
        $Z_1^-\rightarrow$ & (\si{\giga\hertz}) & (\si{\giga\hertz}) & $+1/2\rightarrow$ & (\si{\giga\hertz}) & & polarization & (\si{\giga\hertz\per\tesla}) & (\si{\giga\hertz\per\tesla}) \\
        \hline
        
        $R_1^-$         & $340926.1^\ddag$  & 340925.0   & $-3/2$    & $\star$  & $\pi$ & $\pi$      & $\star$ & 2.2 \\ 
        $R_1^+$         & n/a          & 340941.8   & $+3/2$    & n/a          & n/a & $\sigma$        & n/a         & 19.1 \\
        $\tilde{R}_1^-$ & 340964.9  & 340951.9   & $-3/2$    & 3.73         & $\pi$ &       & 13.9        & 30.2 \\
        $\tilde{R}_1^+$ & 340976.7  & 340967.5   & $+3/2$    & 4.34         & $\pi$ &       & 18.5        & 47.2 \\
        \hline
        $R_2^-$         & $341079.7^\ddag$  & 341078.5   & $-1/2$    & 3.86         & $\sigma$ & $\sigma$   & 5.5         & 7.0 \\
        $R_2^+$         & $341084.2^\ddag$  & 341083.0   & $+1/2$    & 2.95         & $\sigma$ & none   & 14.9        & 14.4 \\
        $\tilde{R}_2^-$ & $341106.3^\ddag$  & 341105.1   & $-1/2$    & 1.8          & $\pi$ &       & 37.6        & 35.0 \\
        $\tilde{R}_2^+$ & $341110.5^\ddag$  & 341109.3   & $+1/2$    & 2.64         & $\pi$ &       & 44.0        & 42.4 \\
        \hline
    \end{tabular}
    \\\raggedright$\star$ Properties not resolved due to high optical depth. $^\dagger$Zero-field data; values have relative precision of \qty{0.2}{\GHz} based on the reference cavity FSR, but have a global offset uncertainty of $\pm\qty{3}{\GHz}$.  ${}^\ddag$Transitions included in the `fine' fit.
    \label{CombinedTable}
\end{table*}

\subsection{Geometric assignment}

    At fields above \SI{2.50}{\tesla}, the crystal has a phase transition into the PM phase. The spectra simplify significantly, because there is no longer diversity in the orientation of the spins.
    
\begin{figure}
        \centering
        \includegraphics[width=0.95\columnwidth]{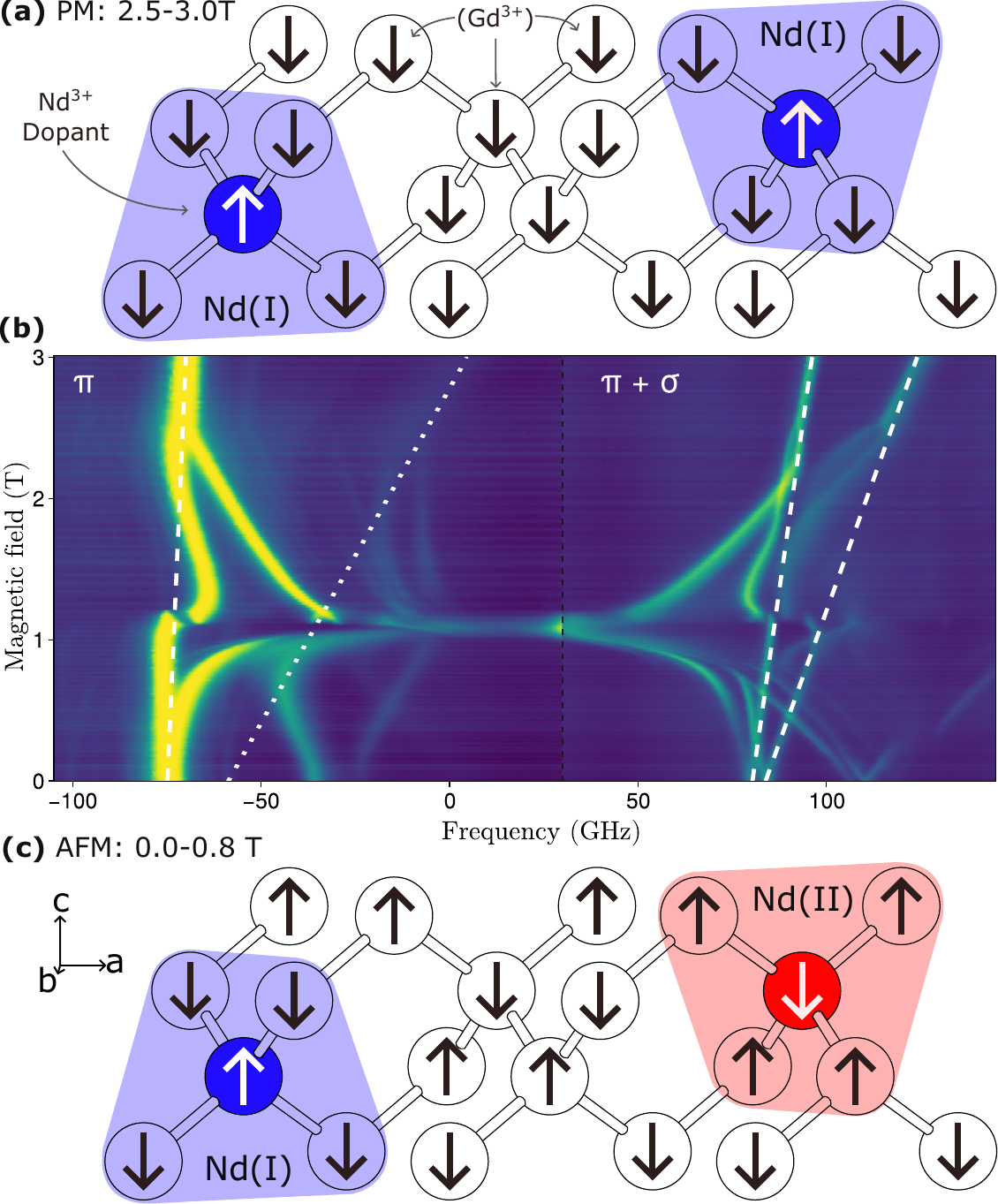}
        \caption{
            Because of the negative $g$-factors of the Nd$^{3+}$ ions, the local spin environment of all the Nd$^{3+}$ in the paramagnetic phase is identical to one of the environments in the antiferromagnetic phase. This can be seen by extrapolating spectral lines between phases.
            (a) Ground-state configuration of Nd:GdVO$_4$ in the paramagnetic (PM) phase at high magnetic field (2.5 to \qty{3}{\T}). Arrows denote spins for an upwards-oriented applied field.
            (b) Absorption spectra as a function of applied magnetic field. White dashed guide lines track field-dependent transitions and illustrate the continuous connection between the paramagnetic and antiferromagnetic (AFM) regimes. A weak transition in the PM phase has a plausible $g$-factor to be the fourth transition we do not observe in the AFM phase and it is shown with a white dotted guide line. In the AFM phase, two magnetically nonequivalent Nd$^{3+}$ sites [Nd(I) and Nd(II)] give rise to distinct transitions. The positive field tuning of these transitions enables assignment to the corresponding Nd sublattices. To best show all transitions, the low-frequency data are $\pi$ polarized, whereas high-frequency data is a mix of both polarizations; the separation is marked by a black dashed line.
            (c) Ground-state configuration of Nd:GdVO$_4$ in the AFM phase at low magnetic field.
            In (a,c) the crystal structure is not to scale.
        }
        \label{fig:GeoAssign}
    \end{figure}

    In this phase, three strong transitions dominate the spectra and increase in frequency with applied field. In Fig.~\ref{fig:GeoAssign}(b) this straight line behavior is extrapolated back down to zero field, demonstrating that this extrapolation agrees with a subset of the zero-field absorptions observed in the AFM regime.
    A fourth transition is optically weak in the PM phase and is not resolved at zero field.

    The reason the high-field transitions appear to point to low-field transitions is that the local spin environment of the Nd$^{3+}$ ions is the same in both cases. 
    In the AFM phase, the magnetic structure is dictated by the exchange and anisotropy of the Gd$^{3+}$ ions, giving two magnetic subsites, Nd(I) and Nd(II), shown in Fig.~\ref{fig:GeoAssign}(c).
    In the PM phase, the magnetic structure is dominated by the magnetic moments of the ions. Gd$^{3+}$ ions have a $g$-factor of $g_\mathrm{Gd}\approx 2$, whereas the $g$-factors of the Nd$^{3+}$ ground doublet are expected to be negative \cite{afzeliusEfficientOpticalPumping2010}. The physical meaning of a negative $g$-factor for an electron system is that the spin expectation value $\langle S_i\rangle$ aligns parallel to the magnetic moment $\langle\mu_i\rangle=-\mu_B\langle L_i+2S_i\rangle$. Therefore, the Gd$^{3+}$ spins align antiparallel to the applied field, whereas Nd$^{3+}$ ions align parallel. Therefore, in the PM phase the Nd$^{3+}$ ions and their nearest neighbors order identically to Nd(I) ions in the AFM phase, shown schematically in Fig.~\ref{fig:GeoAssign}(a,c).
    
    Such extrapolation has no physical meaning within the SF phase, but it becomes highly informative when interpreting the AFM regime, where each transition splits into two branches corresponding to the two oppositely oriented sublattices.
    
    From this analysis, the transitions that increase in frequency with increasing field can be unambiguously assigned to the Nd(I) sublattice, which simultaneously fixes the assignment of the negatively tuning branch to the opposing Nd(II) sublattice. The extrapolated PM branches align closely with the strongest observed zero-field transitions in both $\pi$ and $\sigma$ polarizations.
    The transitions 30--40\,GHz higher in frequency than these main lines are then assigned to sidebands where a spin excitation of the host occurs simultaneously with optical excitation of Nd$^{3+}$.

\subsection{Mean-field model}

    \begin{figure*}
        \centering
        \includegraphics[width=0.95\linewidth]{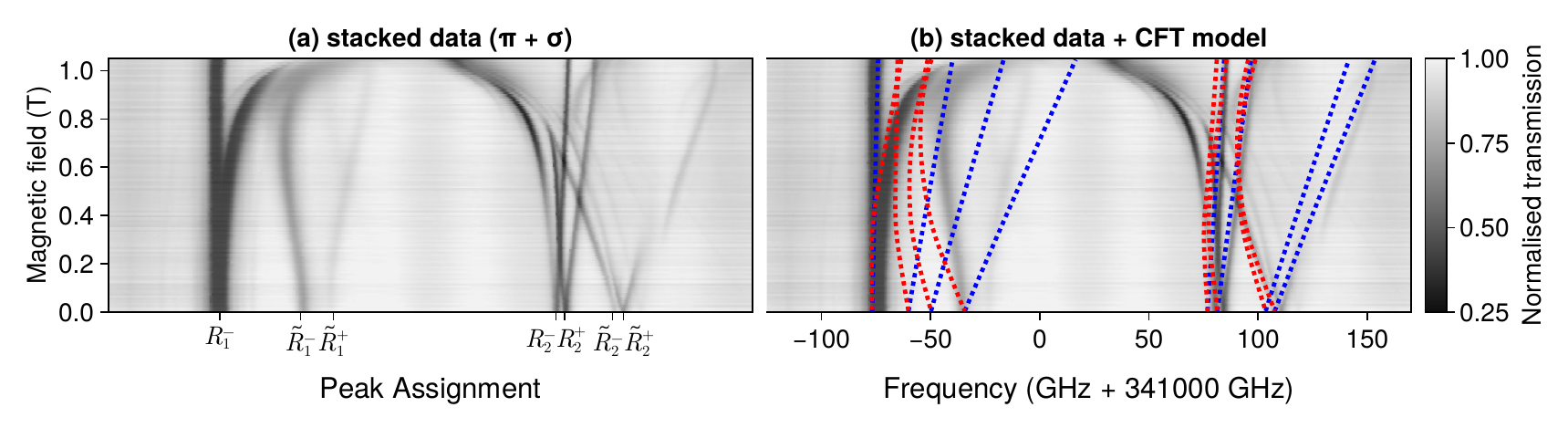}
        \caption{(a) Absorption spectra in the antiferromagnetic (AFM) regime measured for $\sigma$- and $\pi$-polarized light, shown overlaid. (b) The same data with the crystal-field theory (CFT) model overlaid. Colored lines indicate the field-dependent tuning of transitions associated with the blue (Nd(I)) and red (Nd(II)) sublattices.}
        \label{fig:CFTModel}
    \end{figure*}

    With the experimental transitions assigned, the spectra were modelled using a crystal-field Hamiltonian including spin--orbit coupling, Zeeman interaction, and exchange coupling to the ordered Gd sublattices. Figure~\ref{fig:CFTModel}(b) shows the calculated transition frequencies overlaid on the experimental AFM-phase spectra. The two colored branches correspond to transitions associated with the two magnetically nonequivalent antiferromagnetic sublattices, blue corresponding to the Nd(I) site and red to the Nd(II) site. The fitted parameters of the model are given in Tab.~\ref{tab:CF_parameters}. 

    \begin{table}[ht]
    \centering
    \caption{Comparison of crystal-field and free-ion parameters found in this work to initial parameters consisting of crystal-field parameters from \textcite{andersonInterpretiveCrystalfieldParameters1994c} and free-ion parameters from \textcite{carnallSystematicAnalysisSpectra1989}, as well as a crystal-field model for YVO\textsubscript{4} \cite{guillot-noelOpticalSpectraCrystal1998}. Only varied parameters are shown and all others are taken from Ref.~\cite{carnallSystematicAnalysisSpectra1989}. All parameters except for $B_\mathrm{mf}$ have units of \unit{\per\cm}. 
    \textcite{guillot-noelOpticalSpectraCrystal1998} have different signs for $B^4_4$ and $B^6_4$, which together can be understood simply as an alternate coordinate system, rotating the $xy$ plane by \qty{45}{\degree}.}
    \label{tab:CF_parameters}
        \begin{tabular}{c|rrrr}
        \hline
        Parameter & Initial \cite{andersonInterpretiveCrystalfieldParameters1994c,carnallSystematicAnalysisSpectra1989} & This work & Nd:YVO\textsubscript{4} \cite{guillot-noelOpticalSpectraCrystal1998}    \\
            \hline\hline
            $B^2_0$   & $-30.6$  & $-54.8(19)$   & $-200$  \\
            $B^4_0$   & $578$   & $512(17)$   & $628$   \\
            $B^4_4$   & $968$ & $1045(15)$ & $-1136$  \\
            $B^6_0$   & $-1077$ & $-1118(10)$ & $-1233$  \\
            $B^6_4$   & $-234$   & $-194(14)$   & $149$   \\
            $F^2$      & 73018   & $68150(83)$  & $^\dagger71365$    \\
            $F^4$      & 52789   & $55314(81)$  & $^\dagger51993$    \\
            $F^6$      & 35757   & $29515(134)$  & $^\dagger35232$    \\
            $\zeta$    & 885.3   & $880.8(3)$   & $869.1$    \\
            $J$        & -       & $-0.0115(5)$   & -     \\
            $B_\mathrm{mf}$ & -  & \qty{0.942(7)}{\T}  & -     \\
            \hline
        \end{tabular}\\
        ${}^\dagger$Slater parameters calculated from given Racah parameters. 
    \end{table}

    The model reproduces both the zero-field splittings and the observed field dependences across the AFM and PM regimes, providing strong support for the electronic and sublattice assignments derived from the geometric analysis. In particular, the locations of the zero-field transitions and the magnitude of the Zeeman slopes are well captured by the fitted parameters.

    The model predicts two sets of transitions corresponding to two distinct Nd$^{3+}$ environments in the antiferromagnetic (AFM) phase. The first, associated with the sublattice that persists into the paramagnetic phase, exhibits excellent agreement with experiment up to the spin-flop transition. The second, associated with the AFM-exclusive sublattice, displays weaker agreement, particularly at higher fields where the experimental transitions show strong nonlinearity not captured by the present model. This indicates behaviors beyond the assumptions of the fixed mean-field and exchange picture, consistent with a local change in gadolinium orientations around this sublattice.
    
    A quantitative comparison of the principal transition frequencies is given in Table~\ref{CombinedTable}, which lists observed and calculated zero-field frequencies, linewidths, and field sensitivities. The agreement supports the assignment of the $R_{1,2}^{\pm}$ manifolds and some of the cooperative transitions. 
    
    In Tab.~\ref{tab:g-factor}, we compared the calculated $g$-factors and crystal-field quantum numbers of the model to experimental values for Nd:YVO\textsubscript{4}, and the signs and magnitudes are consistent.

    \begin{table*}
        \caption{$g$-factors and quantum numbers derived from the crystal-field model compared to experimental values for YVO\textsubscript{4}. Where appropriate, references for $g$-factor signs are given separately from their magnitudes.
        $\mu$ and $m_J$ are given for the lower state of the doublet assuming a small positive $B_z$. 
        The calculated ground-state wavefunction is $\lvert Z_1,\mu=+1/2\rangle = 0.749\lvert{}^4\text{I}_{9/2},m_J=+1/2\rangle+0.634\lvert{}^4\text{I}_{9/2},m_J=-7/2\rangle+\ldots$, hence it has significant $m_J=+1/2$ and $-7/2$ components. All values are similar to measured values for YVO\textsubscript{4} including their signs. 
        }
        \label{tab:g-factor}
        \centering
        \setlength{\tabcolsep}{8pt}
        \begin{tabular}{cc|cccccc}
            & Doublet & $\mu$ & $m_J$  & $g_\parallel$ & $g_\perp$ & $f_\parallel$ & $f_\perp$  \\ \hline
           Nd:GdVO\textsubscript{4} & ${}^4\text{I}_{9/2}(Z_1)$ & $+1/2$ & $+1/2,-7/2$ & $-1.52$ & $-2.13$ & $+0.79$ & $+0.94$ \\
           calculated & ${}^4\text{F}_{3/2}(R_1)$ & $-3/2$ & $-3/2$ & $-1.21$ & $+0.30$ & $+1.72$ & $+0.29$ \\
           This work & ${}^4\text{F}_{3/2}(R_2)$ & $-1/2$ & $-1/2$ & $+0.52$ & $-0.81$ & $-0.47$ & $+1.17$ \\ \hline
           Nd:YVO\textsubscript{4}& ${}^4\text{I}_{9/2}(Z_1)$ & $+1/2^*$ &  & $-^*\num{0.915\pm0.004}^\dagger$ & $-^*\num{2.361\pm0.003}^\dagger$ \\
           experimental & ${}^4\text{F}_{3/2}(R_1)$ & $-3/2^*$ & & $-^*\num{1.14\pm0.04}^\ddagger$ & $+^\ddagger0.28^\S$ \\
           ${}^*$\cite{afzeliusEfficientOpticalPumping2010}, ${}^\dagger$\cite{goldnerUnderstandingLuminescenceRareearthdoped2007}, ${}^\ddagger$\cite{mehtaOpticalZeemanEffect2000}, ${}^\S$\cite{hastings-simonSpectralHoleburningSpectroscopy2008}. & ${}^4\text{F}_{3/2}(R_2)$ & & & $\pm\num{0.49+-0.09}^\ddagger$ & $\pm\num{0.84+-0.04}^\ddagger$ 
        \end{tabular}
    \end{table*}

\section{Discussion}

The results show that some of the complex behavior of optical dopants in antiferromagnets can have simple, intuitive origins, and that small extensions of crystal-field models to neighboring spins---even with just two parameters for those spins---are very informative about the dopants' behavior.
Such a model reduces the risk of overfitting the data but necessarily cannot explain all behavior in the system.

In particular,
the reduced agreement between measured and calculated transition frequencies at fields above about \qty{0.5}{\T} for the Nd(II) environment is not surprising. 

At low fields, the ordering is dominated by exchange, and the Nd(II) sublattice has its magnetic moment oriented antiparallel to the applied field, in the opposite direction to a free ion.
Approaching the spin-flop transition, a local reconfiguration of the Nd$^{3+}$ ion and its neighboring Gd$^{3+}$ ions may minimize energy, particularly because of our $\sim\qty{10}{\degree}$ angle between applied field and the $c$ axis.
In the future, as intuition is built, a reasonable next step would be to expand the model to the full spin-structure of neighboring gadolinium ions. A challenge here will be the large number of gadolinium spin states; for the four nearest neighbor gadolinium ions, there are $8^4$ possible states, however local structure around the Nd$^{3+}$ ion may extend to next-nearest neighbors or beyond.

The polarizations with which we observe transitions agree well with the electric-dipole selection rules in $D_{2d}$ symmetry, with the exception of the $Z_1^-\rightarrow R_2^+$ transition, which was observed with $\sigma$ polarization. This transition is expected to be dark as the two states share the same crystal field quantum number $\mu=+\frac12$. In fact, the transition from a single initial state to a final Kramers doublet is never expected to be the same polarization for both doublet states, whereas we observe both $R_2^+$ and $R_2^-$ with the same polarization.
Reconciliation would be possible if that transition were a magnetic-dipole transition or if we had misassigned the states and the two $\sigma$-polarized transitions were in fact from different Kramers doublets. Neither of these explanations seems sufficient. We can calculate magnetic-dipole transition strengths from our model and $Z_1^-\rightarrow R_2^+$ is weak for $\sigma$ ($B_z$) polarization; moreover it is about as strong as the $\pi$ ($B_x,B_y$) magnetic-dipole transition strength to $R_1^+$, which is dark. The misassignment possibility is unlikely as it would require the exchange splitting of the $^4\text{F}_{3/2}$ doublets at zero field to be $\sim\qty{150}{\GHz}$, much bigger than the $\approx\qty{20}{\GHz}$ seen in Er:GdVO\textsubscript{4} \cite{hiraishiLongOpticalCoherence2025}; the $g$-factors of the transitions are also opposed to this scenario.

The splitting of the Kramers doublets at zero field due to exchange of $\sim\qty{5}{\giga\hertz}$ ($=zJS_\mathrm{Gd}$) provides a promising microwave transition frequency for microwave--optical transduction, as it is frequency-matched to many superconducting qubit platforms. A challenge for this application, however, is the highly $\pi$-polarized strongest transition, which limits the ability to obtain comparable oscillator strengths for optical pump and signal modes. An applied field perpendicular to the $c$-axis may relax polarization selection rules to provide more equal oscillator strengths between these modes.

\section{Conclusion}

We reported polarization-resolved optical absorption spectroscopy of the ${}^4\text{I}_{9/2}\rightarrow{}^4\text{F}_{3/2}$ transition of Nd$^{3+}$ in antiferromagnetically ordered GdVO$_4$ at \qty{70}{\milli\kelvin} and fields up to \qty{3}{\tesla}. Below the N\'eel temperature the lines are narrow, with linewidths of \qtyrange{2}{5}{\GHz}. As the applied magnetic field is increased, the AFM, spin-flop and paramagnetic phases of the host are clearly visible. Because the neodymium dopants interact both with the host spins---via the exchange interaction---and with an external magnetic field, our results are sensitive to the sign of the magnetic $g$-factors. Our spectra illustrate this nicely: Because the ground state has negative $g_\parallel$, the local spin environment of the neodymium dopants in the high-field paramagnetic phase is identical to one of the AFM orientations, which can be seen by extrapolating the spectra through the spin-flop phase.

This work adds to the comparatively few high-resolution spectroscopic characterisations of rare-earth ions in magnetic host crystals. It is a step towards using these materials both for their potentially long coherence times and for the transduction of quantum signals between the microwave and optical domains.

\section{Acknowledgements}
This work was supported by Quantum Technologies Aotearoa, a research programme of Te Whai Ao –the Dodd-Walls Center, which is funded by the New Zealand Ministry of Business, Innovation and Employment through International Science Partnerships (Contract No. UOO2347).

L.S.T. is supported by the Aotearoa New Zealand Tāwhia te Mana Research Fellowships, administered by the Royal Society Te Apārangi (Contract No. MTP-UOO2502).

\bibliography{refs,refs2}

\end{document}